\documentclass[conference]{IEEEtran}
\usepackage{graphicx}
\usepackage{amsmath}
\usepackage{cite}
\usepackage{xcolor}
\usepackage{soul}
\usepackage{booktabs}
\usepackage{xurl}
\usepackage{algorithm}
\usepackage{algorithmic}   

\usepackage[hidelinks]{hyperref}
\newcommand{\tabfont}{\footnotesize}

\begin{document}

\title{Stratum: A Hardware-Software Co-Design Framework for SNN Inference on RISC-V}

\author{
\IEEEauthorblockN{
Wajahat Riaz, Muhammad Adeel Pasha, Shahid Masud and Muhammad Ali Siddiqi
}
\IEEEauthorblockA{
Department of Electrical Engineering, Lahore University of Management Sciences, Pakistan\\
Email: \{25060008, adeel.pasha, smasud, m.siddiqi\}@lums.edu.pk
}
}

\maketitle
\begin{abstract}
Efficient spiking neural network (SNN) inference on RISC-V processors requires tracing implementation costs back to the compiler and hardware decisions that produce them. Training environments, compilers, and cycle-level simulators, however, typically operate as distinct tools, making this attribution difficult. This paper presents Stratum, a framework that integrates an snnTorch-trained SNN, an MLIR-based compiler stack, and a cycle-level gem5-based RISC-V model into a unified workflow. Stratum exposes two optimization levers, software decisions in the generated code and hardware configuration choices in the simulated core, and uses stage-level cycle measurements to identify performance bottlenecks. We demonstrate Stratum on a two-layer N-MNIST network with three controlled comparisons: (i)~dense vs.\ spike-gated lowering of \texttt{snn.linear}, (ii)~output-outer vs.\ spike-major weight layout with spike-gating enabled in both binaries, and (iii)~L2 cache capacity with both layouts. Our spike-gated lowering reduces simulated inference time by $64\times$ (453.2\,ms to 7.04\,ms), and our spike-major layout cuts it by a further 37.2\% (to 4.42\,ms) at 4\,MiB L2, with the predicted class unchanged. The layout speedup grows from $1.59\times$ to $1.88\times$ when L2 shrinks to 1\,MiB. These results show that Stratum enables rapid early evaluation of design choices at both the software level (compiler stack) and the hardware level (cycle-level gem5 simulation).
\end{abstract}


\section{Introduction}
\label{sec:intro}

Spiking neural networks (SNNs) communicate through sparse binary events (spikes), so most synaptic operations can in principle be skipped. Whether this sparsity actually lowers the inference cost on a processor depends on decisions at two levels: how the compiler translates the network into code, including how its weights are laid out in memory, and how the processor's architectural features, including its memory system, are configured. For SNN inference on RISC-V cores, these decisions interact, so they must be evaluated together before silicon is available, which requires relating measured performance back to the specific compiler and hardware choices behind it.

Prior work covers the individual pieces. Neuromorphic processors such as TrueNorth~\cite{merolla2014} and Loihi~\cite{davies2018}, and accelerators such as SpinalFlow~\cite{narayanan2020}, exploit spike sparsity in dedicated hardware, but many edge deployments run on general-purpose cores. On the software side, snnTorch trains SNNs with gradient-based methods~\cite{eshraghian2023}, the Neuromorphic Intermediate Representation (NIR)~\cite{pedersen2024,nirdocs} exchanges trained models across frameworks, and SNN-MLIR~\cite{gener2026snnmlir} compiles NIR models through MLIR~\cite{lattner2021}, a compiler infrastructure that translates a program step by step from high-level operations into lower-level code. For evaluation, gem5~\cite{binkert2011} models a processor cycle by cycle, and NeuroBench~\cite{yik2025} benchmarks neuromorphic algorithms and systems.

However, these pieces are not connected into a co-design flow. Training frameworks, compilers, and simulators run as separate tools, so the measured cost cannot be easily linked to either a software decision or a hardware choice. SNN-MLIR, the closest compiler, verifies that compiled networks match their reference outputs and run across platforms, but does not measure how its compiler decisions affect cost on a modeled core, and its default translation of synaptic layers computes a dense matrix--vector product that ignores spike sparsity.

This paper presents Stratum\footnote{SpikeDuet is available at \url{https://github.com/chaselab-lums/stratum}.}, a framework for SNN hardware--software co-design that takes a network from snnTorch training to cycle-level simulation on a RISC-V core. Software and hardware decisions are made at compile and simulation time, respectively, so either can be varied while the other is held fixed. Stratum addresses the cost-attribution and sparsity gaps for the fully-connected (FC) feedforward networks that SNN-MLIR supports. The contributions of this paper are:
\begin{itemize}
\item An integrated workflow connecting an snnTorch-trained SNN to MLIR-based compilation and cycle-level RISC-V simulation, with a controlled-comparison protocol and stage-level cycle measurements that attribute cost to individual software and hardware decisions and locate bottlenecks.
\item Two SNN-specific compiler passes (code transformation steps) in \texttt{snn-opt}: spike-gated lowering (translation into explicit loops) of synaptic layers, which skips inactive inputs, and a spike-major weight layout that stores the weights of each input contiguously.
\item An N-MNIST case study (Section~\ref{sec:usecase}) in which spike-gated lowering reduces simulated inference time by $64\times$ and spike-major layout adds a $1.59\times$ speedup that grows to $1.88\times$ as the L2 cache shrinks, showing that a software decision cannot be judged without its hardware context.
\end{itemize}

\section{The Stratum Framework}
\label{sec:framework}

\subsection{Background: NIR and SNN-MLIR}
\label{sec:background}

NIR~\cite{pedersen2024,nirdocs} stores a trained network in a framework-independent form: a directed graph whose nodes are layers, with each neuron model described by its defining differential equations. SNN-MLIR~\cite{gener2026snnmlir} converts this graph into MLIR code expressed with a small set of SNN-specific operations, called a dialect. Each synaptic layer becomes one \texttt{snn.linear} operation, which computes the input currents of a layer's neurons from the incoming spikes and the layer's weights. The spike vector, the weight matrix, and the resulting currents are each held in a memory buffer (\texttt{memref}). A lowering pass then rewrites \texttt{snn.linear} as ordinary nested loops (\texttt{linalg}/\texttt{scf}) that load from and store to these buffers, which the LLVM backend later compiles into machine instructions. Stratum's software levers act at this lowering step: which loops \texttt{snn.linear} is rewritten into, and how the weights are arranged inside their buffer.

\subsection{Architecture}

\begin{figure}[!t]
\centering
\includegraphics[width=\columnwidth]{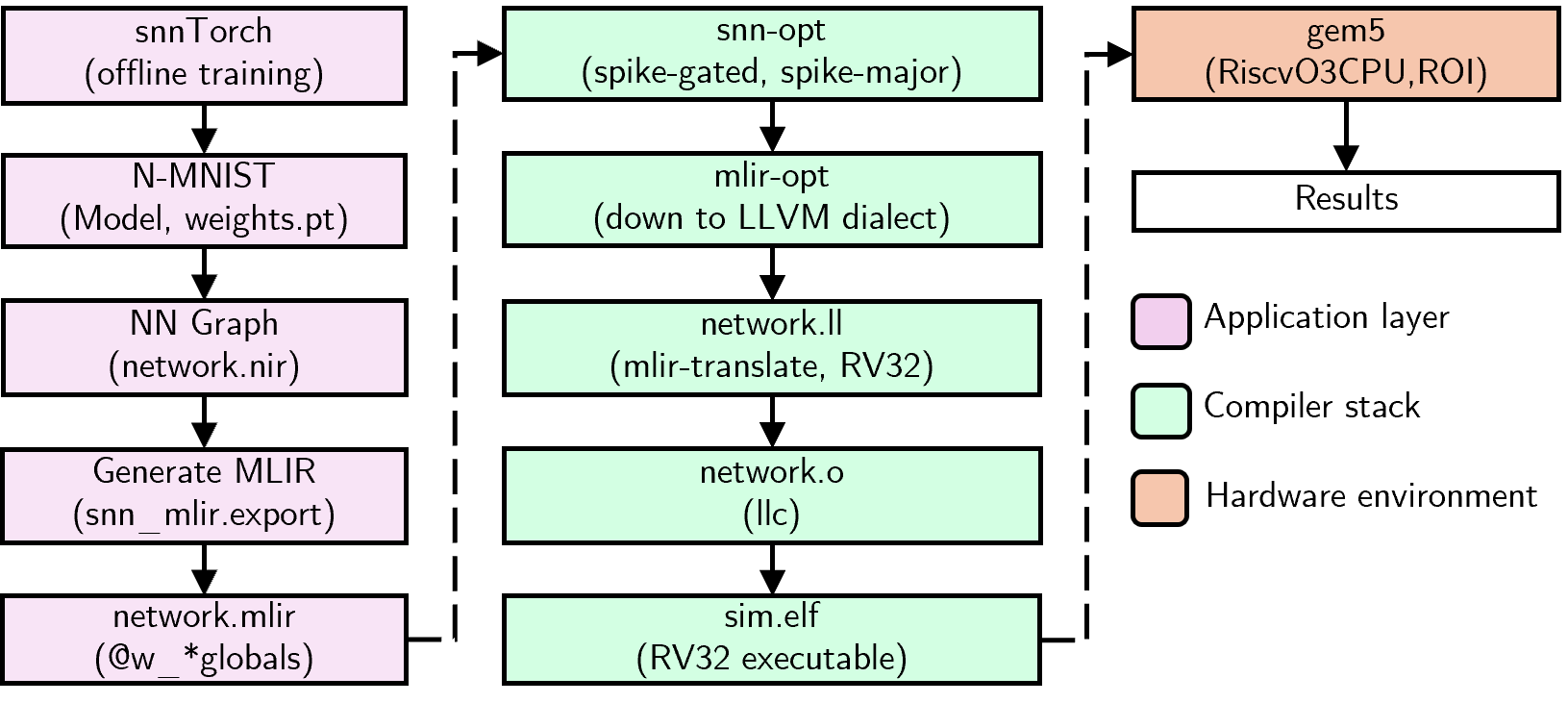}
\caption{Stratum workflow: offline training and export, compiler lowering to \texttt{sim.elf}, and gem5 ROI statistics. }
\label{fig:framework}
\end{figure}

Stratum connects three components, shown in Fig.~\ref{fig:framework}: an SNN application layer, an MLIR-based compiler stack, and a cycle-level hardware simulator. The application layer trains the network with snnTorch on N-MNIST~\cite{orchard2015}, exports it with frozen weights through NIR to SNN-dialect MLIR, and provides the input spike recordings and a host driver.

The compiler stack (\texttt{snn-mlir}) is where software optimization choices are made. Its \texttt{snn-opt} tool runs the lowering passes, including the two that Stratum adds: spike-gated lowering, which skips inactive inputs, and spike-major layout (\texttt{k-major}), which stores the weights of each input contiguously. Standard MLIR passes then lower the result to LLVM IR (\texttt{mlir-translate}), and \texttt{llc} compiles it into a bare-metal RISC-V executable, \texttt{sim.elf}.

The hardware environment (gem5) executes this binary cycle by cycle, modeling instruction execution, caches, and memory stalls, and reports statistics for a defined region of interest (ROI). Software changes, such as a different lowering, produce a new executable, whereas hardware changes, such as a different cache size, only rerun the same executable with new simulator settings. This makes it possible to change one parameter at a time while keeping everything else fixed.

\subsection{Measurement Protocol}

Stratum supports two complementary forms of analysis. First, controlled comparisons measure the effect of a single software or hardware change: two runs differ in exactly one parameter, and the predicted class is checked to be unchanged as a guard against transformations that alter the network's behavior. Second, optional cycle-counter markers (\texttt{rdcycle} reads) delimit individual stages of inference, so their costs can be compared and bottlenecks identified. The markers themselves add execution time, so we report both marked and unmarked cycle counts, together with the share of the ROI the marked stages cover. We also retain the intermediate MLIR, which links each stage's cost to the compiler transformation that produced it.

Three further rules keep the measurements reliable: (i)~each run measures a single input vector, since in our harness the ROI statistics of a multi-vector run describe only the last vector; (ii)~the simulation is deterministic, so rerunning a configuration reproduces its cycle count exactly and a single run per configuration suffices; and (iii)~the configuration behind every reported result is recorded, so any result can be regenerated.

\section{Case Study: N-MNIST Classification}
\label{sec:usecase}

We apply Stratum to an N-MNIST classifier, first describing the network and the simulated RISC-V core, then evaluating two software decisions and one hardware decision.

\subsection{Workload and Setup}

We evaluate Stratum on a two-layer FC SNN trained on N-MNIST~\cite{orchard2015}. Each 1\,ms frame contains 2312 binary inputs ($2\times34\times34$, two polarity planes). The first FC layer (\texttt{fc1}) maps these inputs to 256 hidden neurons, and the second (\texttt{fc2}) maps the hidden layer to 10 output neurons. Each FC layer is followed by a leaky integrate-and-fire (LIF) layer. The LIF decay is 0.5, the threshold is 1, and the reset uses delayed subtraction. Weights are represented as 32-bit floating-point values. The predicted class is the output neuron with the most spikes over 300 time steps. On the first 1{,}280 test samples (ten batches of 128), the model reaches 92.19\% accuracy. Both the input frames and the LIF outputs are binary, so synaptic computation can skip inactive inputs; with 2312 inputs, \texttt{fc1} dominates this cost and is the focus of the optimizations below.

The compiler lowers the model through standard MLIR passes and compiles it with \texttt{llc -O3} for RV32IMACF. The resulting executable is linked as a bare-metal program with a host driver that maintains membrane state and marks the start and end of the ROI. The default gem5 configuration models an out-of-order RISC-V core (\texttt{RiscvO3CPU}) operating at 2\,GHz, with decode, rename, dispatch, and commit widths of 4, an issue width of 8, a 128-entry reorder buffer, 128 integer and 128 floating-point physical registers, and a tournament branch predictor. The L1 instruction and data caches are 32\,KiB, eight-way, with 64-byte lines and 2-cycle tag, data, and response latencies. The L2 cache is 4\,MiB, eight-way, with 64-byte lines, 20-cycle tag, data, and response latencies, and a mostly inclusive policy. No cache warm-up is applied before the region of interest. 

Each ROI covers all 300 forward steps for one input vector, representing 300\,ms of input activity. Outside the ROI, the host performs a one-time vector initialization, copies the selected input sample into the working buffer, and prints the final class. Inside the ROI, each of the 300 steps performs per-step input conversion, one forward step, and output-spike accumulation.

\subsection{Software Optimization Decisions}
\label{sec:software}

To demonstrate software control knobs of Stratum, we examine two software decisions: (i)~spike-gated lowering and (ii)~spike-major weight layout. 


\subsubsection{Spike-Gated Lowering}
\label{sec:sparse-vs-dense}

\begin{figure*}[!t]
    \centering
    \includegraphics[width=\textwidth]{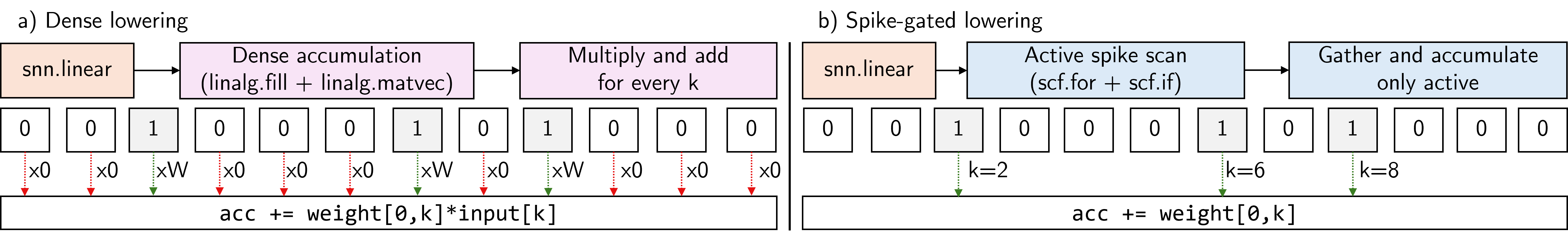}
    \caption{Software optimization decisions evaluated in Stratum.
    (a) Dense lowering and (b) spike-gated lowering of
    \texttt{snn.linear}, with three active inputs in a toy $[N,K]$ layer.}
    \label{fig:decisions}
\end{figure*}

The first decision concerns how the compiler lowers the synaptic computation of an FC layer. For binary input $x_k$, the synaptic current of output neuron $n$ can be written as a sum over the set of active inputs:
\begin{equation}
I_n =
\sum_{k=0}^{K-1} W[n,k]x_k
=
\sum_{k\in\mathcal{S}} W[n,k],
\qquad
\mathcal{S}=\{k:x_k=1\}.
\label{eq:current}
\end{equation}
A conventional lowering of \texttt{snn.linear} produces a dense matrix--vector product that performs a mul-add for each of the $N\cdot K$ input-weight pairs, including inactive inputs, so its work per time step is proportional to $N\cdot K$. Stratum's spike-gated lowering, a new \texttt{snn-opt} pass, instead emits loops that keep the spike test (Fig.~\ref{fig:decisions}): at run time, the generated code first scans the $K$ inputs to collect the active spike indices, then accumulates weights only for those indices. Its work per step is therefore proportional to $K + N\cdot|\mathcal{S}|$, where $|\mathcal{S}|$ is the mean number of active inputs per frame.

For \texttt{fc1}, with $N=256$ and $K=2312$, the dense implementation executes 591{,}872 multiply-accumulate iterations per time step. With about 2.52 active inputs per frame on the first test image, the spike-gated implementation instead executes about $2312 + 256\times2.52 \approx 2{,}957$ scan and accumulate iterations, each cheaper than a multiply-accumulate.
For reference, the dense lowering passes the same functional check and measures 453.176\,ms (906{,}352{,}779 cycles) on the first test image. This is $64\times$ the spike-gated output-outer time and $102\times$ the spike-major time (Section~\ref{sec:layout}). Enabling or disabling \texttt{k-major} does not change this dense timing because the dense loop nest still processes all $N\cdot K$ pairs.

\subsubsection{Spike-Major Weight Layout}
\label{sec:layout}

\begin{figure}[!t]
    \centering
    \includegraphics[width=0.7\columnwidth]{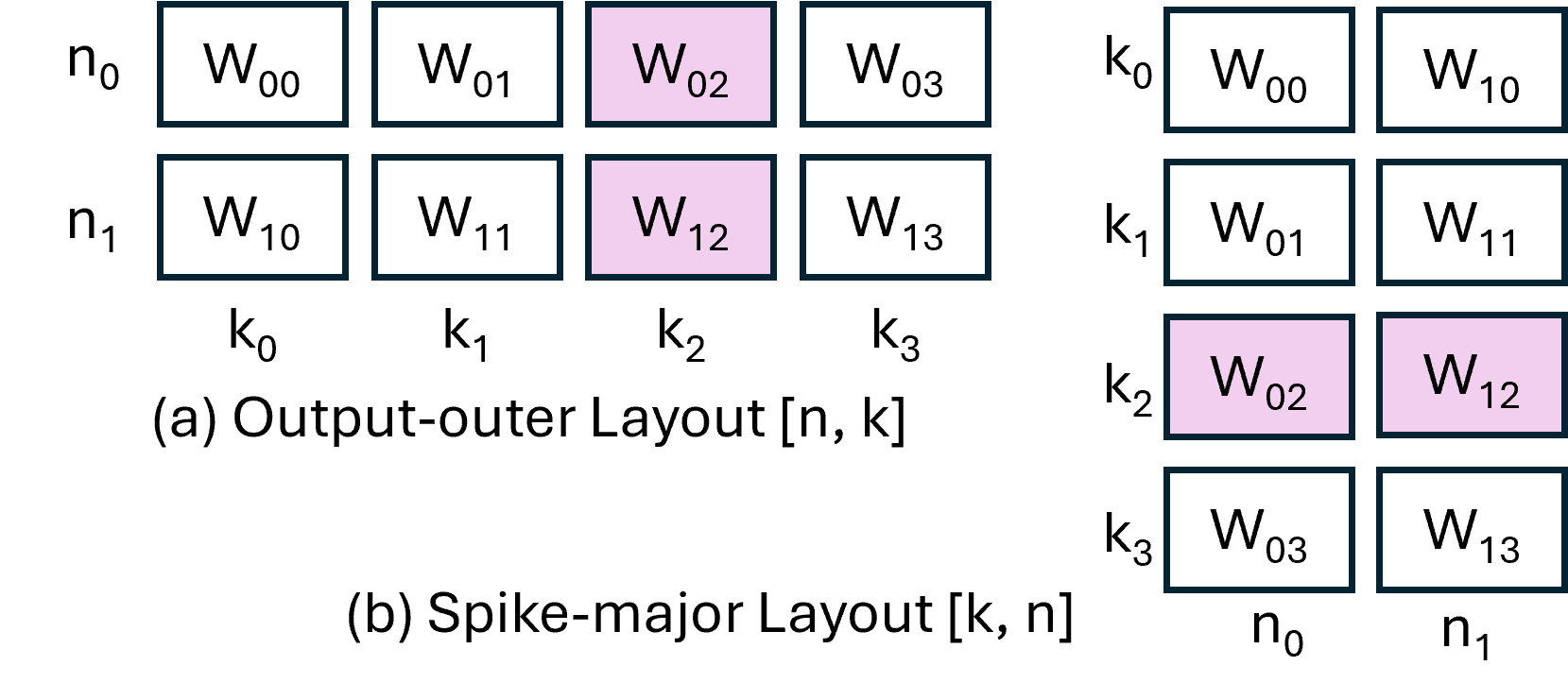}
    \caption{Weight \texttt{memref} layout (toy layer).
    (a)~$[N,K]$ output-outer, (b)~$[K,N]$ spike-major.}
    \label{fig:weight-layout}
\end{figure}

The second software decision concerns how the weights are stored, independent of which synapses are updated. Export uses PyTorch-style $[N,K]$ order, here called \emph{output-outer} (\texttt{k-major=false}), with elements indexed as $W[n,k]$. In row-major storage, stepping to the next output neuron $n$ for a fixed active input $k$ advances the address by $K$ floats; on \texttt{fc1} ($K=2312$), this stride is about 9\,KiB.

The \emph{spike-major} layout (\texttt{k-major=true}) transposes the weights to $[K,N]$, indexed as $W[k,n]$, so all $N$ weights of one active input form one contiguous row (about 1\,KiB for $N=256$), as Fig.~\ref{fig:weight-layout} shows. The transpose is applied once by \texttt{snn-opt} at compile time, so the weights are stored in the binary already in $[K,N]$ order and no transpose runs during inference.

Table~\ref{tab:layout} compares the two layouts with spike-gated lowering in both. On the first test image, output-outer takes 7.038\,ms and spike-major 4.423\,ms to process 300\,ms of input activity, a 37.2\% reduction in simulated inference time, or a $1.59\times$ speedup, with the predicted class unchanged. Across the test images we evaluated, the speedup ranges from $1.55\times$ to $1.69\times$, and both layouts predict the same class in every case.

The speedup decomposes approximately as $1.59\times \approx 1.46\times1.09$: a $1.46\times$ gain in instructions per cycle (IPC) and an 8.1\% reduction in committed instructions. The instruction reduction is consistent with simpler address arithmetic in the contiguous inner loop. Since both layouts perform the same spike-gated computation, the improvement does not come from fewer synaptic updates.

\begin{table}[!t]
\caption{Weight-layout comparison on the first test image at 4\,MiB L2 with spike-gated lowering.}
\label{tab:layout}
\centering
\tabfont
\begin{tabular}{@{}lrrrr@{}}
\toprule
Layout & Time (ms) & Cycles & IPC & Instr. (M) \\
\midrule
Output-outer & 7.038 & 14{,}076{,}427 & 1.157 & 16.29 \\
Spike-major & 4.423 & 8{,}846{,}746 & 1.692 & 14.97 \\
\bottomrule
\end{tabular}
\end{table}

\subsubsection{Stage-Level Measurements}
\label{sec:stage}
We use cycle-counter markers to break the inference cost of the spike-major implementation into stages on the first test image. The marked stages cover 99.75\% of the ROI, but instrumentation raises simulated time from 4.423\,ms (8.85\,M cycles, unmarked) to 4.923\,ms (9.85\,M cycles), an 11.3\% overhead despite only 0.25\% more instructions, consistent with each \texttt{rdcycle} read stalling the out-of-order pipeline. The shares below should therefore be read as approximate.

The active-spike scan is the largest stage, at 38.6\% of marked cycles (3.79\,M), ahead of the gather at 29.3\% (2.88\,M), the first LIF layer at 22.6\% (2.22\,M), and \texttt{fc2} with its LIF layer at 9.5\% (0.94\,M). The scan checks all 2312 inputs at every time step, although only about 2.52 are active. Per step, it therefore runs 2312 loop iterations, against $N\times|\mathcal{S}| = 256\times 2.52\approx 645$ for the gather, or 78\% of their combined iterations. It takes only 57\% of their combined cycles, however, because each gather iteration is more expensive. Since the scan cost depends on the input width rather than on the number of active spikes, much of the inference cost will not shrink as spike activity falls.

\subsection{Hardware Optimization Decision}
We next examine a hardware decision: the capacity of the eight-way L2 cache, swept from 1 to 16\,MiB with everything else held fixed. As Table~\ref{tab:l2} shows, the spike-major implementation is essentially unchanged across the sweep (4.423--4.424\,ms). Output-outer is also flat down to 2\,MiB, but at 1\,MiB it slows to 8.315\,ms, 18\% above its 4\,MiB result, so the layout speedup grows to $1.88\times$.

\begin{table}[!t]
\caption{Simulated inference time vs.\ L2 capacity on the first test image.}
\label{tab:l2}
\centering
\tabfont
\begin{tabular}{@{}lrrr@{}}
\toprule
L2 capacity & Output-outer (ms) & Spike-major (ms) & Speedup \\
\midrule
1\,MiB & 8.315 & 4.424 & $1.88\times$ \\
2\,MiB & 7.046 & 4.423 & $1.59\times$ \\
4\,MiB & 7.038 & 4.423 & $1.59\times$ \\
8\,MiB & 7.038 & 4.423 & $1.59\times$ \\
16\,MiB & 7.038 & 4.423 & $1.59\times$ \\
\bottomrule
\end{tabular}
\end{table}

The complete \texttt{fc1} weight matrix occupies approximately 2.3\,MiB, but the two layouts access it differently. For each active input, output-outer gathers weights from 256 widely separated locations, whereas spike-major reads a contiguous row: roughly 256 separate cache lines against 16, assuming 64-byte lines. Table~\ref{tab:counters} shows the counters collected at 4\,MiB. Spike-major issues far fewer L2 accesses, and although its miss rate is much higher (0.846 against 0.170), it records fewer misses in total, 21{,}629 against 33{,}733, because it touches so few lines. The spike-major layout also reduces L1 data-cache misses from 477{,}611 of 3{,}312{,}199 accesses (14.4\%) to 110{,}352 of 3{,}238{,}481 (3.4\%), and L1D writebacks from 39{,}117 to 540, a $72\times$ reduction.

\begin{table}[!t]
\caption{Memory-system counters at 4\,MiB L2 on the first test image.}
\label{tab:counters}
\centering
\tabfont
\begin{tabular}{@{}lrr@{}}
\toprule
Metric & Output-outer & Spike-major \\
\midrule
L2 accesses & 198{,}088 & 25{,}565 \\
L2 misses & 33{,}733 & 21{,}629 \\
L2 miss rate & 0.170 & 0.846 \\
DRAM reads (bytes) & 2{,}158{,}912 & 1{,}384{,}256 \\
\bottomrule
\end{tabular}
\end{table}

The contiguous access pattern therefore produces fewer total misses and less DRAM traffic, consistent with its insensitivity to cache capacity. Counters at smaller capacities support this reading: as L2 shrinks from 4\,MiB to 256\,KiB, L2 misses grow $3.7\times$ for output-outer (33{,}733 to 124{,}666) but stay nearly constant for spike-major (21{,}629 to 22{,}919). This trend accounts for the output-outer slowdown at 1\,MiB in Table~\ref{tab:l2}, although counters were not collected at that capacity.

\section{Conclusion}
\label{sec:conclusion}

Training environments, compilers, and hardware simulators for spiking neural networks (SNNs) typically operate separately, which makes it difficult to trace implementation costs back to the design choices that cause them. Stratum bridges this gap by connecting SNN training and compilation to cycle-level RISC-V simulation, giving researchers a practical way to explore the hardware-software co-design space for SNN inference and to evaluate design choices early, at both the compiler and the processor-architecture level. In a two-layer N-MNIST workload, spike-gated lowering reduces simulated inference time by $64\times$ relative to dense lowering, and the spike-major weight layout reduces it by a further 37.2\% without changing the predicted class. The layout benefit grows as the L2 capacity is reduced, from $1.59\times$ at 4\,MiB to $1.88\times$ at 1\,MiB, showing that software decisions must be evaluated against the target hardware. Stage-level measurements further show that a substantial share of the inference cost does not scale with the number of active spikes, since the active-spike scan depends on the input width. Future work will use Stratum to derive optimized software-hardware design points from such bottlenecks, and extend it to other NIR neuron models and layer types.

\section*{Acknowledgment}
The authors used Claude (Anthropic) for grammar and text refinement; all technical content, results, and claims were produced and verified by the authors.


\end{document}